\documentclass{IEEEtran}
\usepackage{cite}
\usepackage{amsmath,amssymb,amsfonts}
\usepackage{graphicx}
\usepackage{textcomp,nicefrac}
\usepackage{subfig}
\usepackage{float}
\usepackage{color}
\usepackage{booktabs}
\usepackage{multirow}
\usepackage{graphicx}
\usepackage{enumitem}
\def\BibTeX{{\rm B\kern-.05em{\sc i\kern-.025em b}\kern-.08em
T\kern-.1667em\lower.7ex\hbox{E}\kern-.125emX}}

\begin{document}
\title{Experimental evaluation of neutron-induced errors on a multicore RISC-V platform}
\author{Fernando Fernandes dos Santos, Angeliki Kritikakou, and Olivier Sentieys\\
Univ. Rennes, INRIA, Rennes, France\\
\thanks{}
}

\maketitle

\begin{abstract}
RISC-V architectures have gained importance in the last years due to their flexibility and open-source Instruction Set Architecture (ISA), allowing developers to efficiently adopt RISC-V processors in several domains with a reduced cost. 
For application domains, such as safety-critical and mission-critical, the execution must be reliable as a fault can compromise the system's ability to operate correctly.
However, the application's error rate on RISC-V processors is not significantly evaluated, as it has been done for standard x86 processors.
In this work, we investigate the error rate of a commercial RISC-V ASIC platform, the GAP8, exposed to a neutron beam.
We show that for computing-intensive applications, such as classification Convolutional Neural Networks (CNN), the error rate can be 3.2$\times$ higher than the average error rate.
Additionally, we find that the majority (96.12\%) of the errors on the CNN do not generate misclassifications.
Finally, we also evaluate the events that cause application interruption on GAP8 and show that the major source of incorrect interruptions is application hangs (i.g., due to an infinite loop or a racing condition). 
\end{abstract}

\section{Introduction}
\label{sec:introduction}


Thanks to the open-source Instruction Set Architecture (ISA), RISC-V based processors are today adopted in several domains, including end-user applications~\cite{riscvSmartHomes2020}, High Performance Computing (HPC)~\cite{ficarelli2022RiscVHPC}, and safety-critical applications~\cite{riscvSpace2020, ruospo2019SafetyCritical}.
As RISC-V architecture allows full customization, it enables systems design with reduced cost. Consequently, RISC-V processors became a promising option for safety-critical and mission-critical systems, where power consumption, real-time execution, and reliability are of highest importance. 
However, the majority of RISC-V works mainly focus on design for performance and power consumption, often neglecting reliability. RISC-V architectures need to be evaluated in order to characterize the errors that can reduce their reliability and potentially prevent the system from meeting its mandatory constraints.

The sources of such errors can be environmental perturbations, ionizing radiation, software errors, process, temperature, or voltage variations~\cite{laprie95,nicolaidis99}. It has been demonstrated that the faults caused by radiation have the highest error rates~\cite{Baumann2005}. 
A terrestrial neutron strike may perturb a transistor's state or generate bit-flips in memory or current spikes in logic circuits that, if latched, lead to an error. Neutron-induced events usually are soft because the device is not permanently damaged. 
A new write operation will correctly store the value on the struck memory cell, and consequently, a new operation using the struck logic gate will provide the correct result. 
To evaluate the device reliability under radiation, researchers expose devices such as CPUs~\cite{gabriel2019AMD}, GPUs~\cite{tc2016, luisEntrena2021Protons, dueKojiro2021}, FPGAs~\cite{quinn2005FPGA, Lins2017}, and Tensor Processing Units~\cite{rubens2022TPU} to a radiation flux and measure the error rate. 
Recently, the error rate of soft RISC-V processors synthesized on FPGAs has been measured, when exposed to heavy ions and neutrons~\cite{wilson2019RISCVFPGARad, adria2020SoftRiscvHIons, dilillo2021RiscvNeutron}.
However, as far as we know, none of the existing works considered RISC-V processors physically implemented as an ASIC. 

This work is the first to characterize the error rate of a commercial ASIC RISC-V platform, i.e., the ultra-low-power GAP8 platform from GreenWaves~\cite{gap82018} with a RISC-V master core and a cluster of 8 RISC-V cores, measured on neutron beam experiments. Beam experiments provide a realistic error rate that can be directly scaled to terrestrial flux, and thus, our results can be used to estimate the real error rate of the GAP8 RISC-V platform.
Additionally, to effectively investigate the error rate, we classify the fault outcome of each incorrect execution. We, then, can distinguish, for instance, if the execution hangs or crashes due to a memory error.

We report results from beam testing campaigns that assess the reliability of the GAP8 platform, while running a representative set of codes,
and measure the error rates and the Mean Execution Between Failures (MEBF).
The beam data covers accelerated testing that represents a total of more than 570,000 years of the platform operation. 
The main contributions of this work are as follows:
\begin{itemize}
	\item Provide experimental data and findings on the impact of the neutron-induced errors on GAP8 RISC-V platform.
	\item Reliability evaluation of 5 representative codes from different domains, from signal processing to machine learning. 
	\item Analysis of the fault outcomes for the erroneous executions observed in the experiments using the GAP8 Software Development tool Kit (SDK).
	\item We measure the MEBF to estimate the number of executions that GAP8 devices would perform before experiencing a failure.
\end{itemize}

The remainder of the paper is organized as follows. Section~\ref{sec:background} presents the background on radiation effects on computing devices and the current advances in RISC-V reliability. 
Section~\ref{sec:methodology} presents the tested device, selected codes, and describes the evaluation methodology. The experimental results are presented in Section~\ref{sec:results}, and finally Section~\ref{sec:conclusions} concludes the paper.
\section{Radiation effects on RISC-V devices}
\label{sec:background}

The effects of radiation has been studied for CPUs~\cite{gabriel2019AMD}, GPUs~\cite{tc2016, luisEntrena2021Protons, dueKojiro2021}, FPGAs~\cite{quinn2005FPGA, Lins2017}, and Deep Neural Networks accelerators~\cite{rubens2022TPU}, ARM CPUs~\cite{bodmann2021ARM}, and memories~\cite{naseer2007SRAMRAD, vargas2017SEUMEM, Sulivan2021NVIDIADDR} on beam experiments.  
When the device is exposed to the beam of accelerated particles, faults will be induced on the hardware and may manifest as output errors. 
Beam experiments are the standard methodology for injecting faults on an electronic device. As the chip is equally exposed to the beam of particles the faults will not be limited only to a subset of the device resources.
Consequently, beam experiments provide a realistic device error rate running an application.

As RISC-V processors have been employed in several  application domains, including space exploration~\cite{riscvSpace2020,jan2020Riscvspace}, HPC~\cite{ficarelli2022RiscVHPC}, and safety-critical systems~\cite{ruospo2019SafetyCritical}, it becomes imperative to study RISC-V reliability against neutron-induced errors.
The reliability of soft RISC-V cores synthesized on FPGAs exposed to neutrons and heavy ions has been studied in past works~\cite{wilson2019RISCVFPGARad, adria2020SoftRiscvHIons, dilillo2021RiscvNeutron}. Researchers applied Triple Modular Redundancy for RISC-V and evaluated its reliability in beam experiments~\cite{wilson2019RISCVFPGARad}. In fact, as RISC-V is an open-source ISA, developers can easily modify the architecture to add fault tolerance and evaluate the hardened architecture on beam experiments. 
However, the fault model and error rate measured on beam experiments are significantly hardware-dependent.
As a result, the fault model and error rate generated for soft cores synthesized on FPGAs do not correspond to the fault model and error rate for ASIC processors, and thus,  the evaluation of soft cores synthesized on FPGAs is not representative of applications running on ASIC RISC-V processors.
Note that RISC-V processors built as ASIC are preferable over simulated cores, especially for ultra low power applications, as the use of an FPGA may add an unnecessary power consumption.

To the best of our knowledge, this is the first paper that: (1) evaluates an ASIC commercial mutlicore RISC-V platform (GAP8) on a neutron beam and extracts the realistic error rate for five representative codes; (2) evaluates the observed outcomes on the beam experiments. Furthermore, we keep a trace of each incorrect execution on the beam experiments in order to further analyse it, providing useful insights.

\subsection{Fault propagation on GAP8}

\begin{figure}[htb]%
    \centering
    \includegraphics[width=1\linewidth]{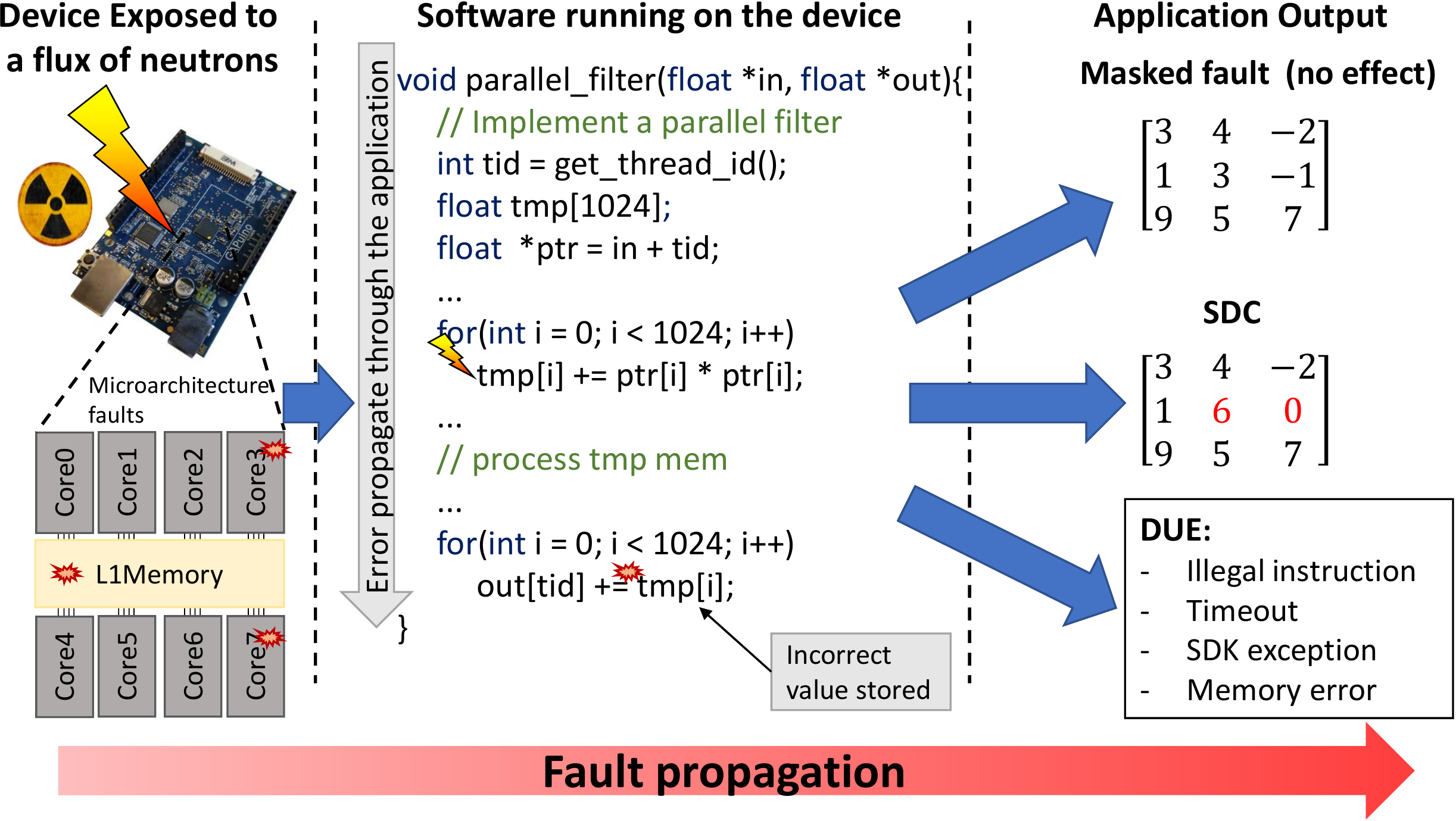}
    \caption{Fault propagation on a multicore RISC-V platform (GAP8) when exposed to a flux of neutrons. The fault is generated at the hardware level as an incorrect functional unit computation on the parallel cores or bit flips in the memory resources. The fault is then propagated on the application execution and manifested as an SDC, DUE, or no observed effect at the application output.}
    \label{fig:background}
\end{figure}

The natural flux of high-energy neutrons at sea level is about 13~$neutrons/(cm^2\times h)$~\cite{Jedec2006}. When iterating with the hardware, high-energy neutrons may generate soft errors in the system. Figure~\ref{fig:background} shows the error propagation on GAP8 ultra-low-power RISC-V platform. The striking particle can generate single or multiple bit flips on a memory resource such as caches, registers, buffers, or even corrupt the value of a functional unit inside a processing core. If the value is used as part of the algorithm computation, the incorrect value will be propagated by the code running on the device. At the application level, the fault may manifest as the following outcomes:
(1) \textbf{No effect on the program output}: The fault is masked, the program output is not affected or the circuit functionality is not affected. 
(2) \textbf{Silent Data Corruption (SDC)}: The program finishes, but the output is not correct, and no flag nor indication is raised.
(3) \textbf{Detected Unrecoverable Error (DUE)}: The system stops working, forcing it to be rebooted or power cycled. A DUE can result from events such as uncorrectable memory events, crashes, or an error that generates an infinite loop (the program hangs). 

Commercial devices like GAP8 often offer a SDK that allows the developers to profile and trace events on the application running on the device. Researchers have successfully monitored neutron-induced events using the SDK or the operating system for ARM~\cite{lindoso2019ARMTRACERAD}, FPGAs~\cite{boyang2017MonitorFPGA}, and GPUs~\cite{dueKojiro2021, santosTNS2022}. The event tracing helps us understand the sources of the errors and the syndromes caused by the radiation in the experiments. On GAP8, we can classify the DUE observed in each code execution into four main sub-classes:
\begin{itemize}
    \item \textbf{Illegal instruction}: The GAP8 SDK returned a crash code that represents an invalid instruction being executed. Illegal instructions can happen when a neutron-generated error corrupts the instruction's opcode;
    \item \textbf{SDK exception}: The GAP8 SDK is responsible for initializing and configuring the device before executing the application for each iteration. If the SDK cannot pre-set the memory or execute the application on the device, it will throw an exception to the user. When the SDK exception happens multiple times sequentially, we perform a power cycle on the board;
    \item \textbf{Timeout}: A Single Event Effect (SEE) on GAP8 may lead to an infinite loop inside the application. To avoid the application hanging indefinitely, we set a \textit{Timeout} limit to stop the running code and start it again;
    \item \textbf{Memory error}: When the application running on GAP8 cannot allocate memory, the API will generate an error reporting a memory allocation error. 
\end{itemize}

\section{Experimental methodology}
\label{sec:methodology}

In this section, we describe the GAP8 RISC-V processor, the codes we characterize, the metrics adopted, and how they are measured for GAP8.

\subsection{Device under test and evaluated codes}

\textbf{Evaluated platform:}
We consider the GAP8, a multicore RISC-V platform from GreenWaves technologies.
The device under test is built with $55nm$ TSMC 55LP CMOS technology. The SoC comprises a cluster of 8 RISC-V cores connected by a Logarithmic Interconnect crossbar, and a RISC-V main processor to manage the cluster. GAP8 has an L1 memory of 64KB and an L2 memory of 512KB, and both memories are shared between the cores of the cluster. Each core of the cluster operates on a maximum frequency of 175MHz. It supports only integer and fixed-point arithmetic. GAP8 is an ultra low power device, and it can process 22.65GOPS with a power consumption of 96mW~\cite{gap82018}.

\textbf{Evaluated codes:} We chose the five representative codes listed in Table~\ref{tab:bench_info} from signal processing to machine learning domains.
The selected codes vary in their complexity and implementation characteristics. Codes such as \textit{MNIST Convolutional Neural Network (CNN)}, \textit{Matrix Addition}, and \textit{Matrix Multiplication} are highly parallelizable, easily distributed between GAP8 eight cores, and computing-intensive. That is, for those codes, the memory instructions are reduced, and most of the instructions are arithmetic ones.
On the other hand, \textit{Finite Impulse Response (FIR)} and \textit{Bilinear Resize} also have computing intensive parts but also have a high number of memory instructions and synchronizations between the main core and cluster. 

The choice of diverse codes increases the quality of the obtained results, which are extendable to different applications~\cite{errorBarHeather2014}.
We select a CNN to classify handwritten digits as a representative case of embedded machine learning algorithms. The CNN is a quantized (16-bit integer) version of LeNet CNN~\cite{lenet98} and has two convolutional layers, each one followed by a ReLU and a MaxPooling layer. The last layer that performs classification is a Linear layer. For the CNN error evaluation, we separate the errors observed on the classification output by their criticality. That is, the errors that are observed on the evaluated CNN can be separated into \emph{Tolerable SDCs} (i.e., the ones that do not affect the inference result) and \emph{Critical SDCs} (i.e., the ones that modify the classification result). 

\begin{table}[]
\centering
\caption{Used codes characteristics. $^*$The execution time is extracted using on the GAP8 SDK.}
\label{tab:bench_info}
\begin{tabular}{lcccc}
\hline
\textbf{}                                                                  & \textbf{Domain}                                             & \textbf{Complexity} & \textbf{\begin{tabular}[c]{@{}c@{}}Used \\ input\end{tabular}}  & \textbf{\begin{tabular}[c]{@{}c@{}}Exec.\\ time$^*$ {[}s{]}\end{tabular}} \\ \hline
\textbf{\begin{tabular}[c]{@{}l@{}}Finite Impulse\\ Response\end{tabular}} & \begin{tabular}[c]{@{}c@{}}Signal\\ processing\end{tabular} & $O(n)$              & \begin{tabular}[c]{@{}c@{}}32KB\\ array\end{tabular}            & 6.4                                                                    \\ \hline
\textbf{\begin{tabular}[c]{@{}l@{}}Matrix \\ Addition\end{tabular}}        & \begin{tabular}[c]{@{}c@{}}Linear \\ algebra\end{tabular}   & $O(n^2)$            & 1KB array                                                       & 3.3                                                                    \\ \hline
\textbf{Bilinear Resize}                                                   & \begin{tabular}[c]{@{}c@{}}Image\\ processing\end{tabular}  & $O(n^2)$            & \begin{tabular}[c]{@{}c@{}}322x242\\ Matrix\end{tabular}        & 9.1                                                                    \\ \hline
\textbf{Matrix Mul}                                                        & \begin{tabular}[c]{@{}c@{}}Linear \\ algebra\end{tabular}   & $O(n^3)$            & \begin{tabular}[c]{@{}c@{}}M1(70x200)\\ M2(150x70)\end{tabular} & 3.7                                                                    \\ \hline
\textbf{MNIST (CNN)}                                                       & \begin{tabular}[c]{@{}c@{}}Machine\\ learning\end{tabular}  & $O(l * n^3)$        & \begin{tabular}[c]{@{}c@{}}MNIST\\ frame\end{tabular}           & 4.9                                                                    \\ \hline
\end{tabular}
\end{table}

\begin{figure}[htb]%
	\centering
	\includegraphics[width=1\linewidth]{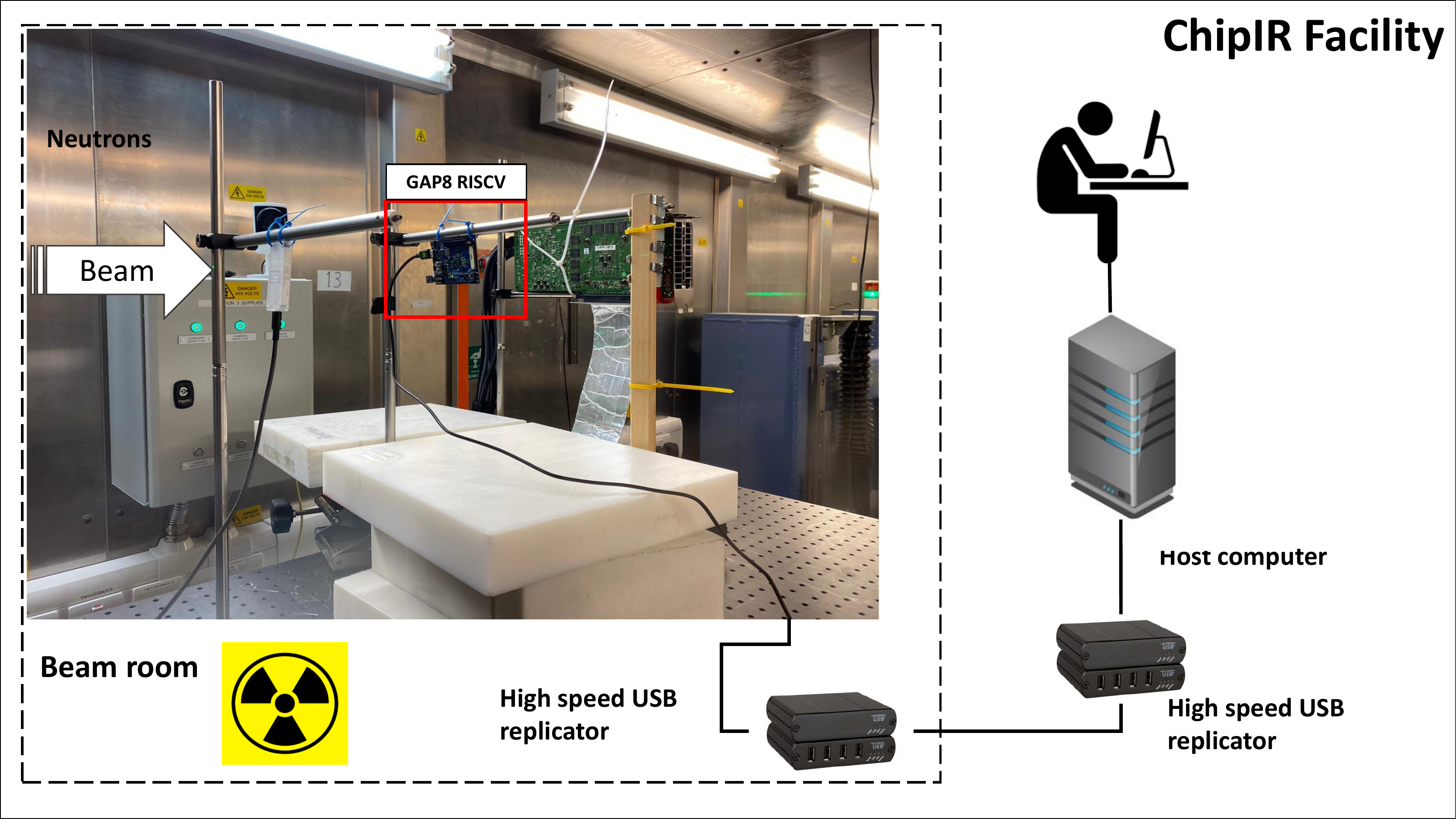}
	\caption{GAP8 setup for the neutron beam experiments.}
	\label{fig:setup_beam}
\end{figure}

\subsection{Beam Experiment Setup}
\begin{figure*}[ht]%
    \centering
    \subfloat[GAP8 Cross section]{
        \includegraphics[width=0.45\textwidth,keepaspectratio]{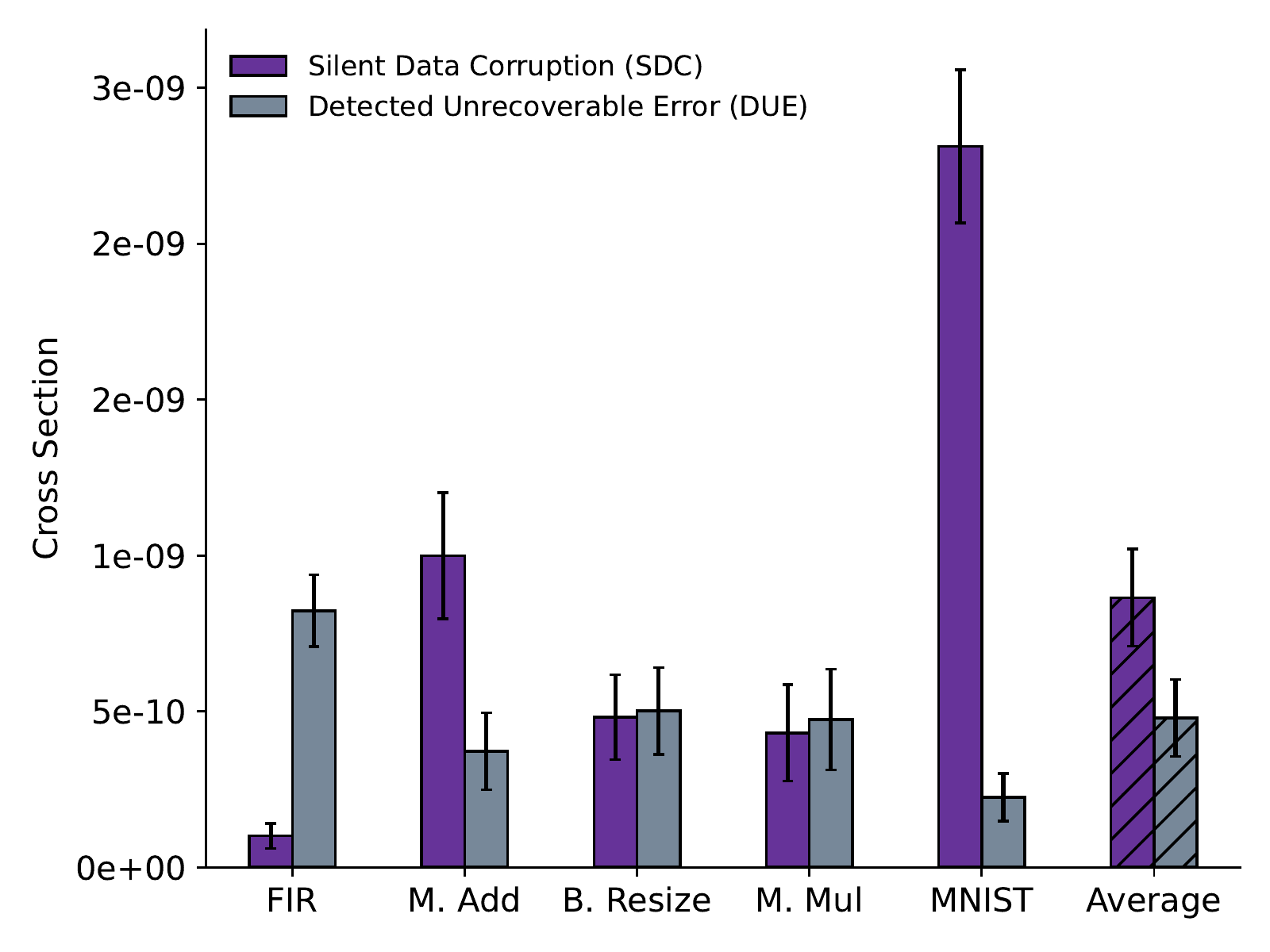}
        \label{fig:gap8_cross_section}
    }%
    \subfloat[GAP8 Mean Executions Between Failure (MEBF)]{
        \includegraphics[width=0.45\textwidth,keepaspectratio]{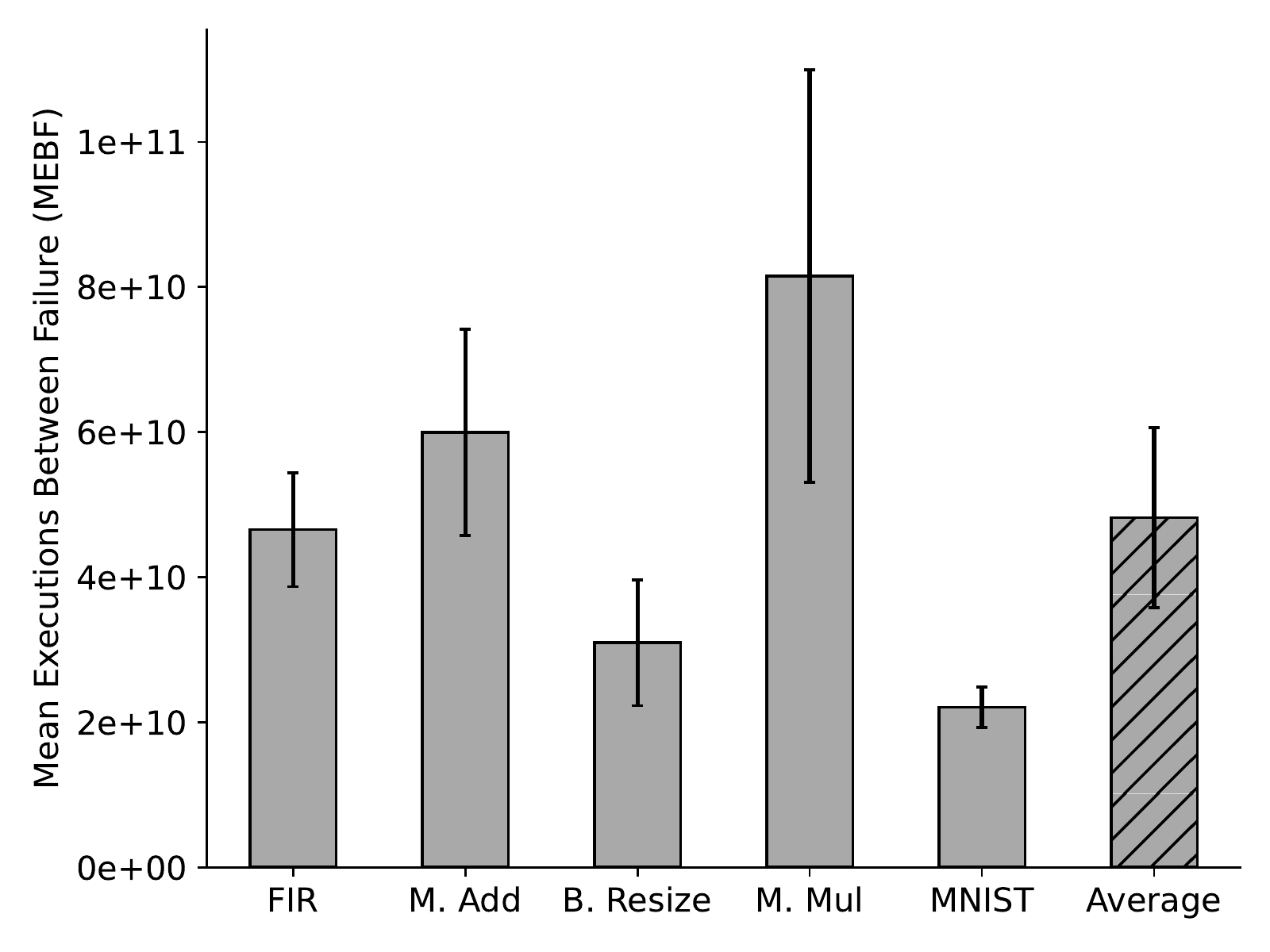}
        \label{fig:mebf}
    }%
    \caption{Figures~\ref{fig:gap8_cross_section} and~\ref{fig:mebf} shows the cross-sections and the Mean Executions Between Failures (MEBF) extracted from the neutron beam experiments for each evaluated code, respectively. The hatched bars represent the average cross-section and MEBF, respectively. It is worth noting that the MEBF is calculated using the execution time provided by GAP8 SDK.}%
    \label{fig:experiments_results}
\end{figure*}
Using beam experiments, we can effectively evaluate the reliability of the GAP8 platform. We expose the GAP8 to a neutron beam to evaluate its error rate. We use the default GAP8 SDK to build and run the experiments. Our experiments are performed at the ChipIR facility of the Rutherford Appleton Laboratory, UK. Figure~\ref{fig:setup_beam} shows the setup of the experiments. The facility delivers a beam of neutrons with a spectrum of energies that resembles the atmospheric neutron one~\cite{Cazzaniga_2018}. The available neutron flux was about $3.5\times10^6 n/(cm^{2}/s)$, $\sim$8 orders of magnitude higher than the terrestrial flux at sea level~\cite{Jedec2006}. 
Since the terrestrial neutron flux is low, it is improbable to see more than a single corruption during program execution in a realistic application. We have carefully designed the experiments to maintain this property (observed error rates were lower than 1 error per 2,000 executions). 
Experimental data can then be used to estimate the error rate in the terrestrial radioactive environment.

We created a software watchdog that runs on the GAP8's host computer (see Figure~\ref{fig:setup_beam}) to monitor and perform recovery from device hangs. 
The software watchdog controls the application under test by executing the applications and monitoring if the execution does not return a result inside a time limit. Then, the program is killed and relaunched if it stops responding in a predefined time interval. We set each code timeout individually depending on the code execution time, up to 6$\times$ the expected execution time. 
The software watchdog also logs if the application crashes during the execution of the code. When the application hangs, the software watchdog uses an ethernet-controlled switch to perform a power cycle on the USB replicator, rebooting the GAP8 board. The board power cycle is necessary to detect when the GAP8 stops responding.

From beam experiments, we can calculate the \textit{cross-section} by dividing the number of observed errors by the received particles fluence $\eta$ ($neutrons/cm^{2}$), as show in Equation~\ref{eq:cs}.

\begin{equation}
\sigma [cm^2] = \frac{\#errors}{\eta}
\label{eq:cs}
\end{equation}    

The fluence $\eta$ is obtained by multiplying the average neutron flux provided by the test facility ($neutrons/(cm^{2} \cdot s$) by the effective execution time. The cross-section ($cm^2$) represents the circuit area that will generate an output error (SDC or DUE) if hit by a particle. The higher the number of computation resources, the higher the cross-section, and the higher the probability of an impinging particle generating an error.

As present in previous works, the cross-section alone does not contain information on the execution time~\cite{Reis2005, rech2014}. Thus, to better evaluate the reliability of tested device and codes, we need to correlate error rates with performance. Hence, we evaluate the Mean Executions Between Failures (MEBF) rate for each tested code. The MEBF is defined as the number of correct executions completed before experiencing a failure~\cite{rech2014}. We calculate the MEBF by dividing the Mean Time Between Failures (MTBF) rate by the code's execution time. The MTBF is the inverse of the cross-section multiplied by the terrestrial flux ($1 / (\sigma \times flux_{terrestrial})$). The MEBF rate is then directly proportional to the tested setup resilience. A higher MEBF rate means more results could be computed without experiencing errors. 

\section{Beam experiments results}
\label{sec:results}

In this section, we present the results of the first evaluation of a commercial parallel ultra-low-power processing processor based on the RISC-V architecture.
We compare the cross-section of the various evaluated codes and the Mean Execution Between Failures (MEBF). 
Additionally, we discuss the erroneous execution outcomes observed in the beam experiments for each code.

\subsection{GAP8 error rate}
Figure~\ref{fig:gap8_cross_section} shows the experimental cross-section measured in the beam experiments (y-axis) for the five codes we evaluated and the average cross-section. 
We present the error rate split into two categories, Silent Data Corruption (SDC) and Detected Unrecoverable Error (DUE).

\begin{figure*}[t]%
    \centering
    \subfloat[]{
        \includegraphics[width=0.45\textwidth,keepaspectratio]{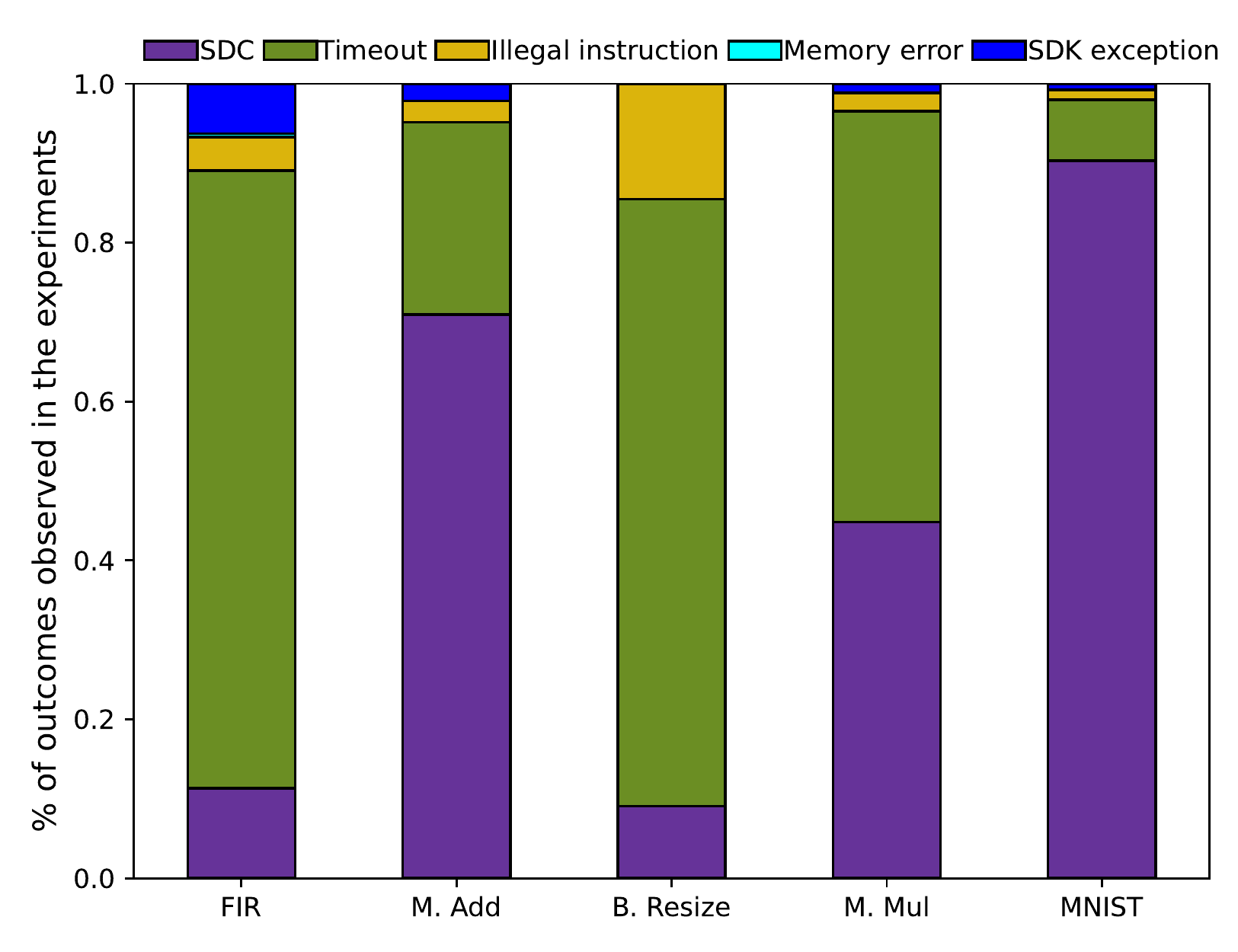}
        \label{fig:outcome_pct}
    }%
    \subfloat[]{
        \includegraphics[width=0.45\textwidth,keepaspectratio]{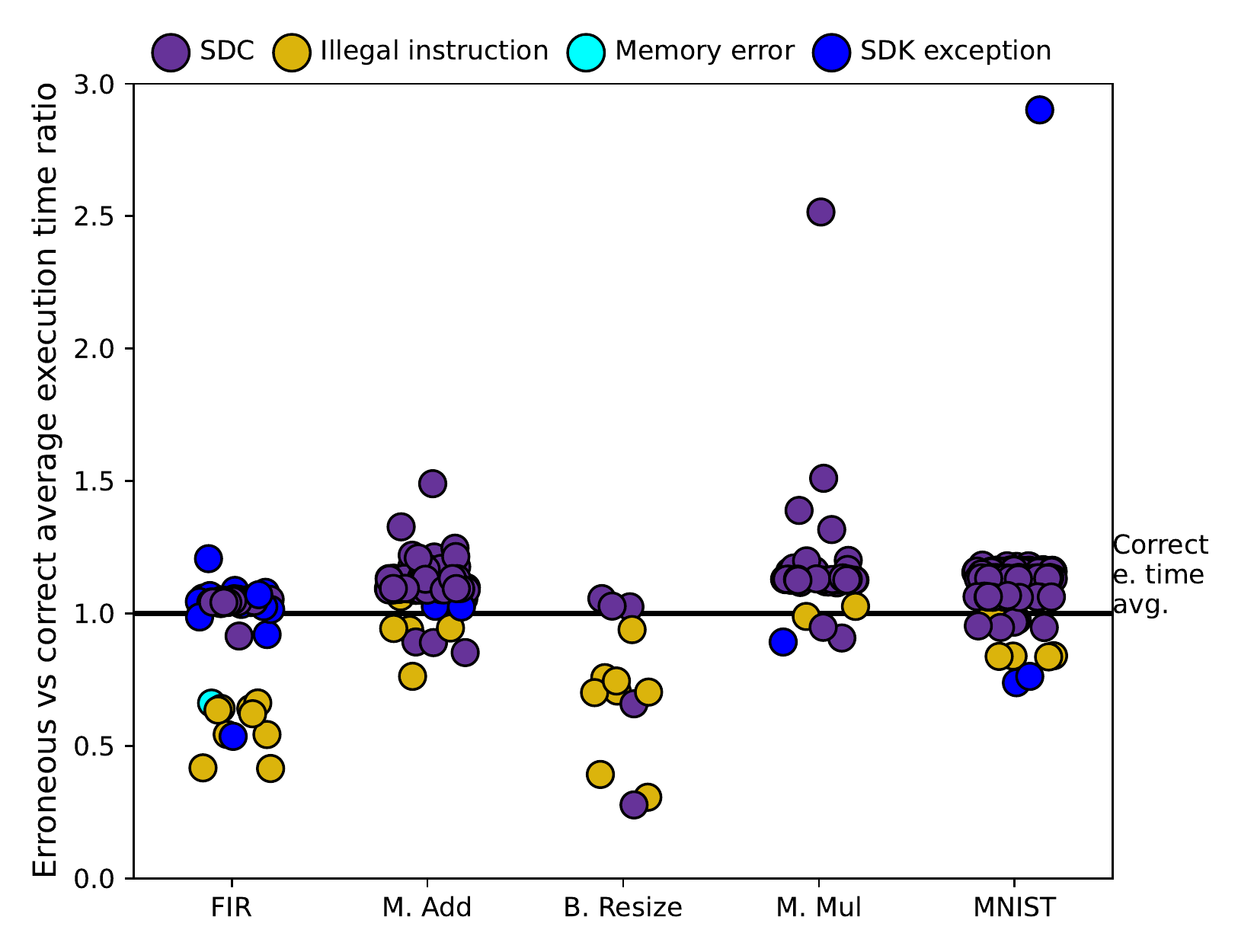}
        \label{fig:time_ratio}
    }%
    \caption{Figure~\ref{fig:outcome_pct} shows the erroneous executions outcomes percentages. Figure~\ref{fig:time_ratio} shows the execution time ratio between the faulty and the fault-free execution time. Timeout executions are not shown as they are always higher than $3\times{}$ the expected execution time. The execution time is measured considering the SDK device setting step.}%
    \label{fig:outcome_analysis}
\end{figure*}

The results show that the DUE error rate varies less than the SDC rate for the evaluated configurations. 
The highest DUE rate variations come from FIR (72\% higher than the average) and MNIST (54\% lower than the average). 
The FIR code is split into steps interleaved with operations sequentially performed by the master core.
The majority of FIR execution is composed of sequential code and managing memory on the main core, leading to less efficient utilization of GAP8 resources. 
Consequently, an incorrect synchronization between the cluster of cores and the main core controller is expected to generate a DUE. 
On the contrary, MNIST CNN is highly optimized to extract the maximum performance of the cluster of cores and reduces any unnecessary management performed by the main core controller and reduce the Direct Memory Access (DMA) operations, leading to a lower DUE rate than the other codes.  
FIR, Bilinear Resize, and Matrix Mul operate over large arrays, which leads to more DMA requests and synchronization actions performed by the main core, increasing the DUE rate.



Figure~\ref{fig:gap8_cross_section} shows that the SDC rate is directly related to how the evaluated codes use the functional units. 
As most of the GAP8 resources are underutilized during FIR execution, the SDC rate is expected to be much lower than the average ($4.5\times$ lower than the average SDC rate).
The Matrix Addition performs the summation of  two linear arrays, which allows the code to be highly parallelized and equally distributed among the cores of the cluster. An error in a core's functional unit or an unprotected memory resource is likely to generate an SDC.
On the contrary, Bilinear Resize and Matrix Multiplication operations can also stress the GAP8 computing units. However, to perform the calculations on large memory arrays that occupy or exceed the L1 memory (See Table~\ref{tab:bench_info}), a high number of memory operations is required, which reduces the usage of the functional units compared to the other codes. As a result, the SDC rate is reduced. 

The MNIST CNN stresses all the computational and memory resources of the GAP8 platform, leading to the highest observed error rate ($3.2\times$ the average SDC rate).
When used for image classification, CNN produces a tensor of values representing the probabilities of classified objects in the frame. Then, after the last layer of the CNN, the probability values will be ranked, and the highest value will be selected.
Although a fault can propagate to the last layer of the CNN, it may not change the classification. 

To evaluate the criticality of the SDCs on MNIST CNN we separate the SDC into \textit{Tolerable SDCs} (a SDC that does not change the classification) and \textit{Critical SDCs} (a SDC that changes the classification). In our experiments, 96.12\% of the SDCs would be classified as \emph{Tolerable SDCs}. Our findings are aligned with previously published works that measured the criticality of the errors on GPUs and TPUs~\cite{karthik2017RangeDet, santosTRCNN2019, rubens2022TPU}. In fact, it has been demonstrated that CNNs can support variations on the output values and still produce the correct classification~\cite{santosTRCNN2019, karthik2017RangeDet}. However, Critical SDCs are a non-neglectable threat, especially for safety-critical and mission-critical systems. Therefore, such errors should be identified and analyzed in order to design hardened techniques for low-power devices, like the GAP8 platform, to at least detect critical errors that may undermine the system's reliability.

\subsection{Mean Executions Between Failures}

The cross-section is a metric that considers the probability of error given the resources used by the application. 
However, the cross-section does not give us any insight into how the performance can impact the reliability. 
Thus, we must use the MEBF to analyze the application's reliability and performance impact. 
Figure~\ref{fig:mebf} shows the MEBF for the evaluated codes. It is worth noting that we calculated the MEBF using the sum of the SDC and DUE cross-section for this evaluation. Then, we can evaluate the MEBF for all the radiation induced-events on the GAP8 platform.

Regarding the MEBF, the differences between the evaluated codes change compared to the cross-section evaluation. MNIST CNN has the highest error rate and, consequentially, even if the code does not have the highest execution time, it has the lowest MEBF ($2.2x10^{10}$ MEBF) among the executed codes. Contrarily, Matrix Multiplication has neither the highest error rate nor highest execution time, leading to the highest MEBF ($8.2x10^{10}$ MEBF). 
It is worth noting that even if Bilinear Resize does not have a high cross-section, it has an MEBF that is only 40\% higher than MNIST due to the high execution time. 
On the other hand, FIR and Matrix Addition have an MEBF near the average.

The cross-section and the MEBF can be directly used to estimate the device's failure rate using the obtained cross-section multiplied by terrestrial flux (13 $neutrons/(cm^{2} \times h$)). If we consider the average cross-section from Figure~\ref{fig:gap8_cross_section}, the sum of SDC and DUE cross-sections ($\approx1.34x10^{-9}$) would yield a failure after each $\approx6.66x10^7$ hours of operation. However, this analysis is valid only when one operating device is considered. The scenario changes when we extend the analysis to multiple devices working in parallel. When considering a million GAP8 based-systems executing simultaneously, the interval of faults would be reduced to 66.6 hours, i.e.,  one fault after each $\approx$2.8 days. This is an alarming result, since many IoT applications on smart home appliances are expected to be in charge of controlling critical systems, such as heating and cooling systems and high voltage mechanisms.

\subsection{Analysis of the incorrect execution outcomes}

Thanks to the setup presented in Section~\ref{sec:methodology}, we can distinguish the outcomes of each incorrect execution (i.e., executions that do not finish and/or produce an incorrect output). Figure~\ref{fig:outcome_pct} shows the percentages of each outcome for each evaluated code in the vertical axis, i.e., SDC, Timeout, Illegal Instruction, Memory error, and SDK exception.

The executions that finished with SDCs or Timeout are the majority of the incorrect executions observed in the experiments. On average, the sum of SDCs and Timeout represents 92\% of the incorrect executions. 
In fact, the probability of a loop control variable, stored in a memory or a register, is modified and thus leads to a Timeout execution, is higher for codes with more instructions such as branch, memory, and synchronization between main and cluster cores, such as FIR and Bilinear Resize. 
Furthermore, for FIR and Bilinear Resize codes, more errors from Illegal Instructions, Memory errors, and SDK exceptions are observed compared to codes composed of more arithmetic instructions, Matrix Add, Matrix Mul, and MNIST CNN.

Additionally, in Figure~\ref{fig:time_ratio}, we report the ratio between the expected average execution time (without faults) and the execution time when an incorrect execution is observed (vertical axis). We draw a black line crossing the y-axis at 1 to represent each code's error-free average execution time. 
The most exciting result is that most executions finishing with an SDC have an execution time near the expected average, but some executions are far from the expected execution time. In fact, even if the SDCs are only errors observed at the application's output, the neutron-induced events may perturb the control flow or increase/reduce the loop iterations, generating a mismatch between the expected output and the observed one. Additionally, the algorithm number of iterations modified by the error may reduce or increase the execution time of the code.



\section{Conclusions}
\label{sec:conclusions}

In this paper, we have evaluated the realistic error rate of the GAP8 RISC-V multicore processor implemented as an ASIC exposed to a flux of neutrons. We have considered five representative codes for our experiments, including a quantized Convolutional Neural Network. 
The error rate of the device extracted from the beam experiments is directly related to how the code uses the processor resources. That is, codes that have to perform more synchronization actions between the main core and the cluster of core and more memory operations, like FIR, have a higher DUE rate. Contrarily, more compute-intensive codes that stress more the functional units and have fewer memory instructions, like the MNIST CNN, have a higher SDC rate. 

Using GAP8 SDK, we have been able to trace the outcomes and the execution time of the incorrect executions observed in the radiation experiments. The results show that a fault can reduce or increase the execution time of an instance of a code that finished with SDC. These results can be used to tune the building and deployment of smart cities, HPC, and safety-critical applications.

Even if the beam experiments are the most realistic way to evaluate an electronic device, they still lack fault propagation visibility. 
In the future, we will inject faults in GAP8 microarchitecture to simulate and evaluate fault propagation. Thanks to RISC-V open-source ISA, we can perform realistic and exhaustive fault simulations in different levels of abstraction, such as RTL models and cycle-accurate simulators. Then, the evaluation of the fault propagation will be an essential step in proposing fault tolerance for parallel multicore RISC-V processors.


\section*{Acknowledgment}

This project has received funding from the European Union’s Horizon 2020 research and innovation program under the Marie Sklodowska-Curie grant agreement No 899546 and with the support of the Brittany Region, and partially funded by ANR FASY (ANR-21-CE25-0008-01) and ANR RE-TRUSTING (ANR-21-CE24-0015-02).
Neutron beam time was provided by ChipIR (DOI:10.5286/ISIS.E.RB2200004-1). 
We acknowledge the researchers that supported and helped with the neutron experiments, Dr. Paolo Rech, Dr. Christopher Frost, and Dr. Carlo Cazzaniga.

\bibliographystyle{IEEEtran}
\bibliography{IEEEabrv,referencespaper}

\end{document}